\documentclass[showpacs,twocolumn]{revtex4}%
\usepackage{amsfonts}
\usepackage{amsmath}
\usepackage{amssymb}
\usepackage{graphicx}%
\begin{document}
\title{Cloud of strings for radiating black holes in Lovelock gravity}
\author{Sushant G. Ghosh $^{a,\;b,}$} \email{sghosh2@jmi.ac.in, sgghosh@gmail.com}
\author{Sunil D. Maharaj $^{a,}$}\email{maharaj@ukzn.ac.za}
\affiliation{$^{a}$ Astrophysics and Cosmology
Research Unit, School of Mathematics, Statistics and Computer Science, University of
KwaZulu-Natal, Private Bag 54001, Durban 4000, South Africa}
\affiliation{$^{b}$ Centre for Theoretical Physics, Jamia Millia
Islamia, New Delhi 110025, India}
\date{\today}
\begin{abstract}
We present exact spherically symmetric null dust solutions in the
third order Lovelock gravity with a string cloud background in
arbitrary $N$ dimensions,. This represents radiating black holes and
generalizes  the well known Vaidya solution to Lovelock gravity with
a string cloud in the background.  We also  discuss the energy
conditions and horizon structures, and  explicitly bring out the
effect of the string clouds  on the horizon structure of  black hole
solutions for the higher dimensional general relativity and
Einstein-Gauss-Bonnet theories.  It turns out that the presence of
the coupling constant of the Gauss-Bonnet terms and/or background
string clouds completely changes the structure of the  horizon and
this may lead to a naked singularity. We recover known spherically
symmetric radiating models as well as static black holes in the
appropriate limits.
\end{abstract}

\pacs{04.20.Jb, 04.70.Bw, 04.40.Nr, 0.4.70.Dy}

\maketitle

\section{INTRODUCTION}
As demanded by  string theory and various higher dimensional scenarios, black holes in $N(>4)$-dimensional spacetimes must have higher curvature corrections.
Among the various gravity theories with higher curvature correction
terms, the so-called Lovelock gravity \cite{dll} is quite special. Lovelock \cite{dll} found the most general second rank tensor with vanishing
divergence being constructed from the second derivative of the metric.  Hence,
its equations of motion contain the most symmetric
conserved tensor with no more than two derivatives of
the metric, and it has been proven to be free of ghosts \cite{zz,bd}.
Lovelock gravity is one of the higher derivative gravity theories,
a natural generalization of Einstein's general relativity, first introduced
by David Lovelock \cite{dll}. The Lovelock action in $N (\geq
4)$-dimensional spacetime is given by
\begin{align}
\label{action}
I=& \frac{1}{2\kappa_N^2}\int  d^N x\sqrt{-g}\sum_{p=0}^{[N/2]}\alpha_{(p)}{\mathcal L}_{(p)}+I_{ matter},\\
 {\mathcal L}_{(p)}:=&\frac{1}{2^p}\delta^{\mu_1\cdots \mu_p\nu_1\cdots \nu_p}_{\rho_1\cdots \rho_p\sigma_1\cdots \sigma_p}R_{\mu_1\nu_1}^{\phantom{\mu_1}\phantom{\nu_1}\rho_1\sigma_1}\cdots R_{\mu_p\nu_p}^{\phantom{\mu_p}\phantom{\nu_p}\rho_p\sigma_p},
\end{align}
where $\kappa_N := \sqrt{8\pi G_N}$. We assume $\kappa_N^2>0$
without any loss of generality, $\alpha_{(p)}$ is an arbitrary
constant with dimension $({\rm length})^{2(p-1)}$, and
$\mathcal{L}_{(p)}$ is the Euler density of a 2$p$-dimensional
manifold. The $\delta$ symbol denotes a totally anti-symmetric
product of Kronecker deltas, normalized to take values $0$ and $\pm
1$~\cite{dll}, which are defined by
\begin{align}
\delta^{\mu_1\cdots \mu_p}_{\rho_1\cdots \rho_p}:=&p!\; \delta^{\mu_1}_{[\rho_1}\cdots \delta^{\mu_p}_{\rho_p]}.
\end{align}
The quantity $\alpha_{(0)}$ is related to the cosmological constant $\Lambda$ by $\alpha_{(0)}=-2\Lambda$.   Among
these terms, the first term is the cosmological constant term, the second
term is the Einstein general relativity (GR) term, and the third and fourth terms
are the second order Lovelock (Gauss-€"Bonnet) and third
order Lovelock terms, respectively.  In this paper, we restrict ourselves to consider the latter three terms of Lovelock gravity, i.e., we start with GR  without a cosmological constant.

In  Lovelock gravity, the case which attracts most attention and which is most extensively studied is the so-called Einstein-Gauss-Bonnet (EGB) gravity. The EGB gravity is a special case of the Lovelock theory of gravitation, whose existence naturally appears in the low energy effective action of heterotic sting theory \cite{Gross}. The spherically
symmetric static black hole  in the EGB gravity was
 first considered  by Boulware and Deser \cite{bd};  this kind of solution for third order Lovelock gravity has been introduced in \cite{ds}. Exact solutions of the former can be
found in \cite{egb}, and the latter in \cite{ll,som,som1,som2}.  The corresponding
EGB black holes in a string cloud model was considered in  \cite{hr,som1}.   Also, several
nonstatic Vaidya-like spherical radiating black hole solutions have been obtained in EGB gravity
\cite{egb-vaidya,gallo,cai}.

It would be interesting to  consider further nonstatic or radiating generalizations
of static black hole  solutions \cite{egb,ll}. It
is the purpose of this Letter to obtain an exact nonstatic solution
in the second and third order Lovelock theory in  the presence of  null dust with a cloud of strings in the background. We shall present a class of
nonstatic solutions describing the exterior of radiating black
holes with null dust endowed with a cloud of strings, i.e., an exact
Vaidya-like solution in the second and third order Lovelock theory for a string cloud model. The radiating black hole solution in GR was obtained by Vaidya \cite{pc} which is a solution of Einstein's
equations with spherical symmetry for a null fluid, i.e., the radially
propagating radiation source.  The nonstatic Vaidya \cite{pc} geometry offers a more realistic background than static geometries, where all back reaction is ignored. It is possible to model the radiating
star by matching it to the exterior Vaidya spacetime (see
\cite{ww,adsg} for reviews on the Vaidya solution and \cite{sgad} for
the higher-dimensional version).  On the other hand  relativistic strings can be used to describe the large scale anisotropy of the universe \cite{lai}, and also to describe extended objects in general relativity \cite{Le1}, e.g., to model stars \cite{gk}, and black holes \cite{hr}.

In this paper we are concerned with radiating black hole solutions in Lovelock theory. We  discuss how higher order curvature corrections in a string cloud background alter black hole solutions and their qualitative features we know from
our knowledge of  black holes in GR. Particular attention will be given to the simpler Gauss-Bonnet case which exhibits most of the relevant
qualitative features. We obtain $N$-dimensional spherically symmetric radiating black hole solutions in a string cloud background with the three terms of Lovelock gravity which are Einstein or GR, Gauss-Bonnet and third order Lovelock terms, and analyze their properties and horizon structures.
The family of solutions discussed here belongs to a Type II fluid.
However, when the matter field degenerates to a Type
I fluid, we can regain  the static black hole solutions \cite{som}. In
particular, our results in the limit $\alpha \rightarrow 0$ and $\beta
\rightarrow 0$ generate models \emph{vis-$\grave{a}$-vis}  $N$-dimensional relativistic solutions \cite{sgad,ms}.

\section{THE STRING-CLOUD MODEL}
The main aim of this paper is to study black hole solutions in $ N $-dimensional Lovelock gravity coupled to a string cloud and discuss their properties.  Expressed in terms of Eddington coordinates,  the metric of a general spherically symmetric spacetime in $N$-dimensions  \cite{sgad} reads
\begin{equation}\label{megb}
ds^2 = - A(v,r)^2 f(v,r)\;  dv^2
 +  2 \epsilon A(v,r)\; dv\; dr + r^2 (d \Omega_{N-2})^2.
\end{equation}
In the above $d\Omega_{N-2}^2 $ is the metric on the ($ N-2 $)-sphere given by
\begin{align}
(d \Omega_{N-2})^2 & = d \theta^2_{1} + \sin^2({\theta}_1) d \theta^2_{2} + \sin^2({\theta}_1)
\sin^2({\theta}_2)d \theta^2_{3} \nonumber \\ & + \ldots + \left[\left( \prod_{j=1}^{N-2} \sin^2({\theta}_j) \right) d
\theta^2_{N-1} \right].
\end{align}
Here $A(v,r)$ is an arbitrary function. It is useful to introduce
a local mass function $m(v,r)$ defined  by \cite{sgad}
\[f(v,r) = 1 - \frac{2
m(v,r)}{(N-3)r^{(N-3)}}.\]
When $m(v,r) = M(v)$ and $A=1$, the
metric reduces to the $N$-dimensional Vaidya metric \cite{sgad,iv}.
Initially $m(v,r) = m_0$ provides the vacuum
Schwarzschild solution in the region $r> 2m_0>0$. The metric in
$N$-dimensional spherically symmetric spacetime is given in terms of the
advanced time coordinate $v$.
Using the Newman-Penrose null tetrad  formalism  the principal null vectors
$l_\mu,n_\mu$ are of the form \cite{gk}
\begin{equation}
l_\mu=\delta _\mu^v,\qquad n_\mu=-\frac{f}{2}\delta _\mu^v+\delta
_\mu^r,
\end{equation}
where $l_\mu l^\mu=n_\mu n^\mu=0,$ $l_\mu n^\mu=-1$. The metric
(\ref{megb}) admits an orthonormal basis defined by unit vectors
\begin{eqnarray}
\hat{v}_\mu=- f^{1/2}\delta _\mu^v+f^{-1/2}\delta _\mu^r,\qquad \hat{r}%
_\mu=f^{-1/2}\delta _\mu^r,
 \label{ov}
\end{eqnarray}
\begin{align}
&\hat{\theta}_\mu=r\delta _\mu^\theta, \;
\hat{\phi}_{(\theta _1)
\mu} = r \sin \theta_1
\delta_\mu^{\phi(\theta_1)}, \nonumber \\
&\hat{\phi}_{(\theta _2) \mu}  = r \sin \theta_1 \sin \theta_2
\delta_\mu^{\phi(\theta_2)}, \; \nonumber \\
&\hat{\phi}_{(\theta _n) \mu}  = r \sin \theta_1 \sin \theta_2
.\;.\;.\;\sin \theta_n \delta_\mu^{\phi(\theta_n)}.
\end{align}
with $n=N-2$. Here, we are interested in the Vaidya-like radiating
black hole solution for a cloud of strings in Lovelock gravity.
Hence the energy momentum tensor we consider includes both a null
dust and a string fluid, i.e.,  a two-fluid description of radial
strings and outward flowing short-wavelength photons that is null
dust \cite{gk}.  In order to see the two-fluid description we use
the above timelike unit velocity vector $\hat{v}^\mu$ and unit
spacelike vectors $\hat{r}^{\mu},
\hat{\theta}^{\mu},\hat{\phi}_{(\theta_n)}^{\mu} $ such that
\begin{eqnarray}
g_{\mu \nu}&=& \hat{v}_\mu \hat{v}_\nu - \hat{r}_\mu \hat{r}_\nu -
\hat{\theta}_\mu \hat{\theta}_\nu - \hat{\phi}_{(\theta_1)\mu}
\hat{\phi}_{(\theta_1)\nu}\nonumber\\ &&- \hat{\phi}_{(\theta_2)\mu}
\hat{\phi}_{(\theta_2)\nu}-.\;.\;.\;- \hat{\phi}_{(\theta_n)\mu}
\hat{\phi}_{(\theta_n)\nu}.
\end{eqnarray}
Next, we consider the theory of a cloud of strings (see \cite{Le1} for further details).
The Nambu-Goto action of a string evolving in the higher dimensional spacetime is given by
\[    I_{\mathrm{S}} = \int_{\Sigma} \mathcal{L} \;  d\lambda^{0} d\lambda^{1}, \hspace{0.2in} \mathcal{L} = m (\gamma)^{1/2},     \]
where $m$ is a positive constant that  characterizes each string,
$(\lambda^{0}, \lambda^{1})$ is a parametrization of the world sheet
$\Sigma$ with $\lambda^{0}$ and $\lambda^{1}$ being timelike and
spacelike parameters \cite{synge}, and $\gamma$ is  the determinant
of the induced metric on the string world sheet $\Sigma$   given by
\begin{equation}
 \gamma_{a b} = g_{\mu \nu} \frac{\partial x^{\mu}}{\partial \lambda^{a}} \frac{\partial x^{\nu}}{\partial \lambda^{b}},
\end{equation}
where $\gamma $ = det $\gamma_{a b}$.
Associated with the string world-sheet we have the string bivector \cite{Le1,gk}
defined by
\begin{equation}
\Sigma^{\mu \nu}=\epsilon^{AB}
\frac{dx^{\mu}}{d{\lambda}^A}\frac{dx^{\nu}}{d{\lambda}^B},
\label{sb}
\end{equation}
where $\epsilon^{AB}$ is the two-dimensional Levi-Civita symbol.  It is
useful to write the bivector, in terms of the unit vectors, as
\begin{equation}\label{eq:bivector}
\Sigma ^{\mu \nu}=\hat{r}^\mu \hat{v}^\nu -\hat{v}^\mu \hat{r}^\nu,
\end{equation}
and the condition that the world-sheet are timelike, i.e., $\gamma
=\frac{1}{2}\Sigma ^{\mu \nu}\Sigma _{\mu \nu}<0$ implies that only the
$\Sigma ^{ur}$ component is nonzero.  Therefore one obtains
\begin{equation}
\Sigma ^{\mu c}\Sigma _c^\nu=\hat{v}^\mu \hat{v}^\nu -\hat{r}^\mu
\hat{r}^\nu.
\end{equation}
The string energy momentum tensor for a cloud of strings is \cite{Le1,gk}
\begin{equation}
T_{\mu \nu}=\rho \; (\gamma)^{-1/2} \Sigma _\mu ^c\Sigma
_{c\nu}, \label{ems}
\end{equation}
where $\rho$ is the proper density of a string cloud. The quantity $\rho \; (\gamma)^{-1/2} $ is the gauge invariant quantity called the gauge-invariant density. The string is characterized by a surface-forming bivector $\Sigma^{\mu \nu} $, and conditions to be a surface-forming are
\begin{eqnarray} \label{eq:imp}
  & & \Sigma^{\mu [\alpha} \Sigma^{\beta \gamma]} = 0, \nonumber \\ & & \nabla_{\mu} \Sigma^{\mu [\alpha} \Sigma^{\beta \gamma]} = 0,
\end{eqnarray}
where the square brackets indicate anti-symmetrization in the enclosed indices.
  The above equations, in conjunction with Eqs.~(\ref{eq:bivector}) and (\ref{eq:imp}), lead to the useful identity
\begin{equation}
\label{eq:usable}
     \Sigma^{\mu \sigma} \Sigma_{\sigma \tau} \Sigma^{\tau \nu} = \gamma \Sigma^{\nu \mu},
\end{equation}
which will be used in the later calculations.  Further $T^{\mu \nu}{}_{;\mu} =0$ implies that
\begin{equation}
\label{eq:div}
     \partial_{\mu} (\sqrt{-\mathbf{g}} \rho \Sigma^{\mu \sigma}) = 0.
\end{equation}
We are seeking spherically symmetric solutions, which restrict the
density $\rho$, and the bivector $\Sigma_{\mu \nu}$ as function of
$v,\; and\;  r$ only.
 Further the only possible surviving component of the bivector $\Sigma$ is $\Sigma^{vr} =  - \Sigma^{rv}$.  Thus $T^v_v=T^r_r=-\rho \Sigma^{vr}$, and  from Eq.~(\ref{eq:div}), we obtain  $\partial_{r} ({r^n T^v_v}) = 0$ which implies
\begin{equation}
T^v_v = T^r_r = \frac{S(v)}{r^n},
\end{equation}
where $S(v)$ is function of $v$. The energy momentum of the two fluid system is $\mathbf{T}_{\mu \nu} =
T^{(n)}_{\mu \nu} + T_{\mu \nu}$, where $
T_{\mu \nu}^{(n)}=\zeta l_\mu l_\nu.$
It is the null fluid tensor corresponding to the component of the
matter field that moves along the null hypersurfaces
$v=\mbox{const.}$ The effective energy momentum tensor for the two-fluid
system, in terms of the unit vectors, can be cast as:
\begin{eqnarray}\label{emt}
\mathbf{T}_{\mu \nu}&=& \zeta l_\mu l_\nu + \rho \hat{v}_\mu \hat{v}_\nu + p_r
\hat{r}_\mu \hat{r}_\nu .
\end{eqnarray}
Furthermore, the energy momentum tensor considered is of the special
case $T^v_r=0$, which means that from the field equation $G^v_r=0$
we get $A(v,r)=g(v)$.  However, by introducing another null
coordinate $\overline{v} = \int g(v) dv$, we can always set, without
any loss of generality, $A(v,r) = 1$ (see also \cite{sgad}).

\section{THE EINSTEIN LOVELOCK ACTION\ AND\ FIELD\ EQUATIONS}
Lovelock gravity is the most general second order gravity theory
in higher dimensional spacetimes, and it is free of ghosts when
expanding in space \cite{dll}. The Lovelock tensor is nonlinear in the Riemann tensor and differs from the Einstein tensor
only if the spacetime has more than four dimensions ($ 4D $).
The action which describes the third order Lovelock gravity, without a cosmological constant in $N$ dimensions,  reads
\cite{dll}%
\begin{equation}
I_{G}=\frac{1}{2}\int_{\mathcal{M}}dx^{N}\sqrt{-g}\left[  R+\alpha_{2}\mathcal{L}_{GB}+\alpha_{3}\mathcal{L}_{\left(  3\right)
} \right] +\mathbf{I}_{M},
\end{equation}
where
\begin{equation}
\mathcal{L}_{GB}=R_{\mu\nu\gamma\delta}R^{\mu
\nu\gamma\delta}-4R_{\mu\nu}R^{\mu\nu}+R^{2},
\end{equation} is the Gauss-Bonnet (GB)
Lagrangian, and
\begin{align}
\mathcal{L}_{\left(  3\right)  }  &  =2R^{\mu\nu\sigma\kappa}R_{\sigma
\kappa\rho\tau}R_{\quad\mu\nu}^{\rho\tau}+8R_{\quad\sigma\rho}^{\mu\nu
}R_{\quad\nu\tau}^{\sigma\kappa}R_{\quad\mu\kappa}^{\rho\tau}\nonumber\\
&  +24R^{\mu\nu\sigma\kappa}R_{\sigma\kappa\nu\rho}R_{\ \mu}^{\rho}%
+3RR^{\mu\nu\sigma\kappa}R_{\sigma\kappa\mu\nu}\nonumber\\
&  +24R^{\mu\nu\sigma\kappa}R_{\sigma\mu}R_{\kappa\nu}+16R^{\mu\nu}%
R_{\nu\sigma}R_{\ \mu}^{\sigma} \nonumber \\
&  -12RR^{\mu\nu}R_{\mu\nu}+R^{3},
\end{align}
is the third order Lovelock Lagrangian. Here $R,$ $R_{\mu\nu\gamma\delta\text{
}}$ and $R_{\mu\nu}$ are the Ricci scalar, Riemann and Ricci tensors,
respectively. Variation of the action with respect to the spacetime
metric $g_{\mu\nu}$ yields the Einstein-Gauss-Bonnet-Lovelock
(EGBLL) equations%
\begin{equation}\label{llee}
G_{\mu\nu}^{E}+\alpha_{2}G_{\mu\nu}^{GB}+\alpha_{3}G_{\mu\nu}^{\left(
3\right)  }=\mathbf{T}_{\mu\nu},
\end{equation}
where $\alpha_{2}$ and $\alpha_3$ are second and third order Lovelock coefficients, and
$G_{\mu\nu}^{E}$ is the Einstein tensor, while $G_{\mu\nu}^{GB}$ and
$G_{\mu\nu}^{\left(  3\right)  }$ are second and third order Lovelock tensors given explicitly as \cite{zz}
\begin{align}
G_{\mu\nu}^{GB} & =2\left(  -R_{\mu\sigma\kappa\tau}R_{\quad\nu}^{\kappa
\tau\sigma}-2R_{\mu\rho\nu\sigma}R^{\rho\sigma}-2R_{\mu\sigma}R_{\ \nu
}^{\sigma}+RR_{\mu\nu}\right) \nonumber \\ &   -\frac{1}{2}\mathcal{L}_{GB}g_{\mu\nu},
\end{align}%
and
\begin{align}
& G_{\mu\nu}^{\left(  3\right)  }  =-3\left(  4R_{\qquad}^{\tau\rho\sigma\kappa
}R_{\sigma\kappa\lambda\rho}R_{~\nu\tau\mu}^{\lambda}-8R_{\quad\lambda\sigma
}^{\tau\rho}R_{\quad\tau\mu}^{\sigma\kappa}R_{~\nu\rho\kappa}^{\lambda}\right. \nonumber
\\
 &   +2R_{\nu}^{\ \tau\sigma\kappa}R_{\sigma\kappa\lambda\rho}R_{\quad\tau\mu
}^{\lambda\rho}-R_{\qquad}^{\tau\rho\sigma\kappa}R_{\sigma\kappa\tau\rho
}R_{\nu\mu}+8R_{\ \nu\sigma\rho}^{\tau}R_{\quad\tau\mu}^{\sigma\kappa
}R_{\ \kappa}^{\rho} \nonumber\\
 & +8R_{\ \nu\tau\kappa}^{\sigma}R_{\quad\sigma\mu}^{\tau\rho}R_{\ \rho}^{\kappa
}+4R_{\nu}^{\ \tau\sigma\kappa}R_{\sigma\kappa\mu\rho}R_{\ \tau}^{\rho
}-4R_{\nu}^{\ \tau\sigma\kappa}R_{\sigma\kappa\tau\rho}R_{\ \mu}^{\rho
}\nonumber\\
& +4R_{\qquad}^{\tau\rho\sigma\kappa}R_{\sigma\kappa\tau\mu}R_{\nu\rho}%
+2RR_{\nu}^{\ \kappa\tau\rho}R_{\tau\rho\kappa\mu}+8R_{\ \nu\mu\rho}^{\tau
}R_{\ \sigma}^{\rho}R_{\ \tau}^{\rho}\nonumber\\
& -8R_{\ \nu\tau\rho}^{\sigma}R_{\ \sigma}^{\tau}R_{\ \mu}^{\rho}-8R_{\quad
\sigma\mu}^{\tau\rho}R_{\ \tau}^{\sigma}R_{\nu\rho}-4RR_{\ \nu\mu\rho}^{\tau
}R_{\ \tau}^{\rho}\nonumber\\
&  +4R_{\quad}^{\tau\rho}R_{\rho\tau}R_{\nu\mu}-8R_{\ \nu}^{\tau}R_{\tau\rho
}R_{\ \mu}^{\rho}+4RR_{\nu\rho}R_{\ \mu}^{\rho}
\left.  -R^{2}R_{\nu\mu}\right)  \nonumber \\  & -\frac{1}{2}\mathcal{L}_{\left(  3\right)
}g_{\mu\nu}.\nonumber
\end{align}
The constants $\alpha_2$ and $\alpha_3$ will help us  track the changes in the equations, when we compare with the
corresponding equations of GR.   We are interested in the exact solutions
of black holes in Lovelock theory coupled to
a string cloud which will be shown to be
generalizations of the earlier solutions. To achieve this,
we need solve the EGBLL equations (\ref{llee}) for the spherically metric \textit{ansatz} (\ref{megb}). In this case only diagonal and one off-diagonal $(r,v)$ EGBLL equations survive.  It is enough to solve the $(v, v)$ component of the EGBLL field equations (\ref{llee}),
which amounts to the equation
\begin{align}\label{f-egb}
 &- \frac{  \left( N-2 \right) }{2r^2} \Big[  \left( 1+2\,{\frac {\alpha\,
 \psi(v,r) }{{r}^{2}}}+{\frac {\beta\,
  \psi(v,r)^{2}}{{r}^{4}}} \right) r \psi'(v,r)  \nonumber \\ &  - \psi(v,r) \Big( (N-3)+{\frac { \left( N-5 \right)
\alpha\, \psi(v,r) }{{r}^{2}}}+ \nonumber \\ &  {
\frac { \left( N-7 \right) \beta\,  \psi(v,r)^{2}}{3{r}^{4}}} \Big) \Big] = T^v_v, \nonumber  \\ & \mbox{and the $(r,v)$ equation takes the form} \nonumber  \\
& - \frac{ \left( N-2 \right)}{2r}  \left( 1+2\,{\frac {\alpha\,\psi(v,r) }{{r}^{2}}}+{\frac {\beta\, \psi(v,r)^{2}}{{r}^{4}}} \right) \dot{\psi} \left( v,r \right) = T^r_v,
\end{align}
in which a dot and a prime denote, respectively, derivatives with respect to $v$ and $r.$ In the above the functions $\psi(v,r)=1-f(v,r)$, ${\alpha}=$ $\left(N-3\right)  \left(N-4\right)  \alpha_{2}$ and
${\beta}=$ $\left(N-3\right)  \ldots  \left(N-6\right)  \alpha_{3}$.
In three and four dimensions Lovelock theory coincides with Einstein general relativity, e.g., for $N=4$, we get%
\begin{equation}
- rf^{\prime}\left(v, r\right) + 1  - f\left( v, r\right)  =S(v),
\end{equation}
which clearly is $\alpha$ and $\beta$ independent, and therefore it will be the
Einstein equation in 4D admitting the solution
\begin{equation}\label{sol4d}
f(v,r) = 1 - \frac{2 M(v)}{r} - S(v),
\end{equation}
where the function $M(v)$ arises due to integration and this is the famous Vaidya solution with a cloud of strings in the background \cite{sdn},  and goes over to Letelier model \cite{Le1} when $M(v)$ and $ S(v) $ are constants. The above solution can be identified as radiating black hole spacetime associated with a spherical mass $M(v)$ centered at the origin of the system of coordinates, surrounded by a spherical cloud of strings.

In higher dimensions ($ HD $), the Lovelock gravities  are actually different. In fact, for $N > 4$ Einstein gravity can be thought of as a particular case of Lovelock gravity since the Einstein-Hilbert term is one of several terms that
constitute the Lovelock action. Hence, for  $N > 4$ and $\alpha = \beta = 0 $, we obtain
\begin{equation}\label{eehd}
(N-2) r^{(N-4)} \left[r f'(v,r) - (N-3)\left(1 - f(v,r)\right)\right] = 2 S(v),
\end{equation}
which admits a solution
\begin{equation}\label{ndsol}
f\left(v,  r\right)  =1-\frac{2M(v)}{(N-3)r^{N-3}}- \frac{2S(v)}{(N-2)r^{N-4}}.
\end{equation}
  We also observed that the above four dimensional solution (\ref{sol4d})  is recovered in the limit $N \rightarrow 4$.  From the $(r,v)$ component of (\ref{f-egb}), we obtain the energy
  density of the fluid as
  \begin{eqnarray}\label{ndensity}
  \zeta(v,r)= \frac{(N-2)}{(N-3) r^{N-2}} \frac{dM}{dv} +  \frac{1}{r^{(N-3)}}\frac{dS}{dv}. \label{density}
  \end{eqnarray}
 It may be noted that the $N$-dimensional solution (\ref{ndsol}) outlined here contains the
 $N$-dimensional version of, for instance, the Vaidya metric \cite{sgad,iv} (when $S(v)=0$). The solution (\ref{ndsol}) can be identified as the $ HD $ version of the Vaidya solution with clouds of strings in the background.   In particular, in the 4D case, the solution (\ref{ndsol}) reduces to (\ref{sol4d}).  The static black hole solutions, in both HD \cite{gallo} and in 4D  \cite{Le1}, can be recovered by setting $M(v) = M, \; S(v) = S$, with $ M $ and $ S $ being constants, in which case the matter is Type I.

 In order to study the general structure of the solution given in (\ref{ndsol}),  we look for the essential singularity.  It is seen that the Kretschmann scalar  for the metric (\ref{megb}) reduces to
  \begin{eqnarray}
  \mathcal{K} & = &  R_{abcd} R^{abcd}= f{''}(v,r)^2 +2(N-2)\frac{f'(v,r)}{r^2} \nonumber \\ & & + 2 (N-2) (N-3)  \frac{f(v,r)^2}{r^4} ,  \label{density}
  \end{eqnarray}
  which on inserting (\ref{ndsol}) becomes
  \begin{align}\label{ks}
   \mathcal{K} &= \frac{4 (N-2)^2(N-1)(N-3)}{r^{2N-2}} M^2(v)  \nonumber \\ & +  \frac{8(N-3)^2(N-2)}{r^{2N-3}} M(v) S(v) \nonumber \\ & +
   \frac{(N^4-12N^3+55N^2-144N+92)}{(N-2)^2 r^{2N-4}} S^2(v),
  \end{align}
  which clearly diverges as $r \rightarrow 0$ indicating the scalar polynomial or essential singularity at $r=0$.   In the higher dimensional case, a fact which deserves to be emphasized is that the cloud of strings alone ($M(v)=0$), unlike in $4D$,  can have an apparent horizon located at $r_{EH}= (2S(v)/(N-2))^{1/(N-4)}$. Thus we have extended the  Letelier \cite{Le1} solutions to nonstatic  higher dimensional spacetimes.

\section{THE\ EGB CASE ($\alpha_2\neq0$ and $\alpha_3=0 $)}
The simplest case in  Lovelock theory arises when we choose $\alpha_3=0 $, which is the well-known EGB theory that embodies nontrivial dynamics for the gravitational field in five-(or higher) dimensional theories.  The static spherically symmetric black hole solutions of EGB theory were first obtained by Boulware and Deser \cite{bd}. We wish to find the general
solution of the Einstein equations for the matter field given by
Eq.~(\ref{emt}) for the metric (\ref{megb}).
Eq. (18) with $\alpha_{3}=0$ takes the form%
\begin{align}\label{s-fegb}
 & - \frac{  \left( N-2 \right) }{2r^2} \Big[  \left( 1+2\,{\frac {\alpha\,
 \psi(v,r) }{{r}^{2}}} \right) r \psi'(v,r)  \nonumber \\ &  - \psi(v,r) \Big( N-3+{\frac { \left( N-5 \right)
\alpha\, \left( \psi(v,r) \right) }{{r}^{2}}} ,
 \Big) \Big] = T^v_v
\end{align}
which may be called the EGB equation. This
equation admits a general solution in  arbitrary dimensions $N$ as follows%
\begin{equation}\label{s-feggb}
f_{\pm}\left(v, r\right)  =1+\frac{r^{2}}{2{\alpha}}\left(  1\pm
\sqrt{1+\frac{4{\alpha}M(v)}{(N-2)r^{N-1}} + \frac{8{\alpha}S(v)}{(N-2)r^{N-2}}} \right),
\end{equation}
where $M(v)$ is an arbitrary function of $v$. The
special case in which $M(v)$ is a nonzero constant we call the
GB-Schwarzschild solution \cite{bd}, for which the global structure is
presented in~\cite{tmgb}. There are two families of solutions which correspond to the sign in
front of the square root in Eq.~(\ref{s-feggb}). We call the family
which has the minus (plus) sign the minus- (plus+) branch solution.
From the $(r,v)$ component of (\ref{f-egb}), we obtain the energy
density of the fluid  which is again given by (\ref{density})
for both branches, where the dot denotes the derivative with respect
to $v$. In order for the energy density to be nonnegative, ${\dot
M} \ge 0$ must be satisfied. In the static case $\dot{ M}=\dot {S}=0$, the solution (\ref{s-feggb}) reduces to the solution which was obtained in  \cite{he}, and for $N=5$  to the results in  \cite{som1}. Further, $\dot{ M}={S(v)}=0$, the solution (\ref{s-fegb})  reduces to the EGB black hole, independently discovered by Boulware and
Deser~\cite{bd}.   It may be noted that similar solutions in five dimensions were reported in \cite{som2}. Also a similar kind of solution in arbitrary dimensions,  but in different context, has been discussed in \cite{cai,gallo}. Asymptotically, the minus branch of the solution (\ref{s-feggb}) goes to HD  general relativistic   limit ${
\alpha} \to 0$, and as it is expected the minus-branch solution in this limit looks like Eq.~(\ref{ndsol}).  There is no such limit for the plus-branch solution.   We will restrict our discussion to the most generic situation, i.e., we shall consider the solution with the general relativistic or minus-branch. Finally, when the string cloud background is switched off we recover  the EGB-Vaidya solution \cite{egb-vaidya,dwd} from the Eq.~(\ref{s-feggb}).

To see the asymptotic behavior of the solution (\ref{s-feggb}), we
take  the limit $r \rightarrow \infty$ or $M(v)=S(v)=0$ in
(\ref{s-feggb}), to obtain  \[\lim_{r \rightarrow \infty}
f_{+}\left(v, r\right) =1+\frac{r^{2}}{{\alpha}}, \hspace{0.1in}
\mbox{and} \lim_{r \rightarrow \infty} f_{-}\left(v, r\right) =1,
\] which means the plus+ branch of the solution (\ref{s-feggb}) is asymptotically de
Sitter (Anti-de Sitter) depending on the sign of $\alpha \; (\pm)$
whereas the minus- branch of the solution (\ref{s-feggb}) is
asymptotically flat.  Further, it is seen that the Kretschmann
scalar (\ref{ks}) diverges as $r \rightarrow 0$ indicating a scalar
polynomial singularity.

\subsection{Energy Conditions}
The family of solutions discussed here, in general, belongs to a Type II fluid defined in \cite{he}. When
$M=M(r)$, we have $\mu =0$, and the matter field degenerates to a type I fluid \cite{ww}. In the rest frame
associated with the observer, the energy density of the matter is given by
\begin{equation}
\zeta = T^r_v,\hspace{.1in}\rho = - T^v_v =  T^r_r = \frac{S(v)}{r^{(N-2)}}, \label{energy}
\end{equation}
 and the principal pressures are $P_i =
T^i_i$ (no sum convention).  Therefore $P_r = T^r_r = - \rho$ and
$P_{\theta_i} = 0$.
\noindent \emph{a) The weak energy
conditions}: The energy momentum tensor obeys the inequality $T_{ab}w^a w^b \geq 0$ for any timelike vector, i.e.,
$
\zeta \geq 0,\hspace{0.1 in}\rho \geq 0,\hspace{0.1 in} P_{\theta_1}
= P_{\theta_2}= .\, .\, .\, = P_{\theta_{(N-2)}} \geq 0.
\label{wec}
$
The strong energy condition and the weak energy
conditions, for a Type
II fluid, are identical \cite{ww}.
\noindent {\emph{b) The dominant energy conditions}}: For any timelike vector $w_a$, $T^{ab}w_a w_b \geq 0$,
and $T^{ab}w_a$ is a non-spacelike vector, i.e.,
$
\zeta \geq 0,\hspace{0.1 in}\rho \geq P_{\theta_1}, P_{\theta_2}=
.\, .\, .\, = P_{\theta_{(N-2)}} \geq 0.$
Clearly, $(a)$ is satisfied if $S(v)\leq 0$. However, $\zeta
> 0$ gives the restriction on the choice of the functions $M(v)$
and $S(v)$. From Eq.~(\ref{ndensity}),  we
observe $\zeta > 0$ requires
\begin{equation}
\frac{(N-2)}{(N-3) r^{N-2}} \frac{dM}{dv} +  \frac{1}{r^{(N-3)}}\frac{dS}{dv} > 0.
\end{equation}
This, in general, is satisfied, if
${d M}/{d v}
> 0$  and ${d S}/{d v} > 0$. On the other hand,  the dominant energy conditions hold if $S(v) \leq 0$, and the function
$M(v)$ is subject to the condition ${d M}/{d v}
> 0$  and ${d S}/{d v} > 0$.
\section{RADIATING BLACK HOLE HORIZONS}
In this section, we study the structure and location of the horizons and compare them
with  general relativity  by using the solution obtained in
the previous section.  We consider the minus-branch solution in
order to compare with the general relativistic case. The line element of
the radiating black hole in EGB gravity has the
form (\ref{megb}) with $f(v,r)$ given by Eq.~(\ref{s-fegb}) and  the
energy momentum tensor (\ref{emt}). We can define  a timelike limit surface as locus where $g(\partial_v, \partial_v) = g_{vv} = 0$ where $\partial/\partial{v}$ is a timelike vector and $L = -{dM}/{dv}$, and $M(v)$ is the mass of the black hole. It is known that  the apparent horizon coincides with the timelike limit surface \cite{jy}.

The luminosity due to loss of mass is given by
$L_M = - dM/dv$, $L_M < 1$ \cite{jy}, and due to clouds of string by $L_S = -
dS/dv$, where $L_M, L_S < 1$.  Both  are measured in the  region
where $d/dv$ is timelike. In order to further discuss the physical
nature of our solutions, we need to introduce their kinematical parameters.
We assume $v =$ constant is an in-going null surface with a future-directed null tangent vector $l^a$. Then we define a future-directed null geodesic by  tangent vector $n^a$ such that
\begin{eqnarray}
l_{\mu}l^{\mu} &=& n_{\mu} n^{\mu} = 0, \; ~n_{\mu} l^{\mu} = -1,\; ~l^{\mu}
\;\gamma_{\mu \nu} = 0.
\label{nvdgb}
\end{eqnarray}
The metric at $v=$ constant will be $ (N-2) $-dimensional, say $\gamma_{ab}$, and let the spacetime metric be $g_{\mu \nu}$. Following York \cite{jy,bc,rm,mrm} a null vector decomposition
of the metric (\ref{megb}) is  of the form
\begin{equation}\label{gab}
g_{\mu \nu} = - l_{\mu} n_{\nu} - n_{\nu} l_{\mu} + \gamma_{ab},
\end{equation}
where
\begin{eqnarray}\label{nvgb}
l_{\mu}& &= \delta_{\mu}^v, \: n_{\mu} = \frac{1}{2} f(v,r) \delta_{a}^v +
\delta_a^r, \nonumber
 \\
\gamma_{\mu \nu} && = r^2 \delta_{\mu}^{\theta_1} \delta_{\nu}^{\theta_1} + r^2 \sum\limits_{i=2}^{N-1}
\left[\left( \prod_{j=1}^{i-1} \sin^2({\theta}_j) \right) \right]
\delta_a^{\theta_i} \delta_b^{\theta_i}, \nonumber
\end{eqnarray}
with $f(v, r)$ given by Eq.~(\ref{s-fegb}).  The optical behavior
of the null geodesic congruences is governed by the Raychaudhuri
equation
\begin{equation}\label{regb}
   \frac{d \Theta}{d v} = K \Theta - R_{\mu \nu}l^{\mu}l^{\nu}-(\gamma_c^c)^{-1}
   \Theta^2 - \sigma_{\mu \nu} \sigma^{\mu \nu} + \omega_{\mu \nu}\omega^{\mu \nu},
\end{equation}
with expansion $\Theta$, twist $\omega$, shear $\sigma$, and surface
gravity $K$. Here $R_{\mu \nu}$ is the $N$-dimensional Ricci tensor and $\gamma_c^c$ is the trace projection tensor for null geodesics.
The expansion of the null rays parameterized by $v$ is
given by
\begin{equation}\label{theta}
\Theta = \nabla_{\mu} n^{\mu} - K,
\end{equation}
where the $\nabla$ is the covariant derivative and the surface
gravity is
\begin{equation}\label{sggb}
K = - l^{\mu} n^{\nu} \nabla_{\nu} n_{\mu}.
\end{equation}

\noindent
The apparent horizon is the outermost marginally trapped
surface for the outgoing photons, which can be either null or
spacelike, that is, it can `move' causally or acausally \cite{jy,rm}
Using Eqs.~(\ref{s-fegb}), (\ref{nvgb}) and (\ref{sggb}) we get
\begin{equation}
K = \frac{1}{2} \frac{\partial f}{\partial r}\label{Kgb}.
\end{equation}
Then Eqs.~(\ref{s-fegb}), (\ref{nvgb}),
(\ref{theta}),  and  (\ref{sggb}) yield the expansion parameter \cite{dwd,mrm}
\begin{equation}
\Theta = \frac{(N-2)}{2r} f(v,r)\label{thgb}.
\end{equation}
The apparent horizons are defined as surfaces such that $\Theta \simeq 0$
which implies that $f(v,r)=0$ and which is equivalent to  $g(\partial_v, \partial_v) = g_{vv} = 0$;  hence the two surfaces coincide for the  spherically symmetric case.  However, they may be different for axially symmetric black holes.

The apparent horizon for  $ 4D $ GR is give by $1 - 2 M(v)/r -S(v)=0$, which admits a real root
\begin{equation}
r_{AH}^{4D} = \frac{2M(v)}{1-S(v)}.
\end{equation}
In the limit, $S(v) \rightarrow 0$ then $r_{AH}^{4D} \rightarrow 2 M(v)$ and as $M(v) \rightarrow 0$ no horizon exists.  Thus the cloud of strings alone, in the $4D$ case, does not have a horizon and hence naked singularity at $r=0$.
Next we are going to calculate apparent horizons for the higher dimensional case which, if they exist, are given by zeros of $f(v,r)=0$, i.e.,  we need to look for a solution
\begin{equation}\label{ahhd}
1-\frac{2M(v)}{(N-3)r^{N-3}}-\frac{2S(v)}{(N-2)r^{N-4}}=0.
\end{equation}
In the $ HD $ GR case, a fact which deserves to be emphasized is that the cloud of strings alone, unlike in $4D$,  can have an apparent horizon located at \[r_{AH}^S= \frac{2S(v)}{(N-2)^{1/(N-4)}}.\]
The apparent horizon is located at, e.g. in the 5D case, at $r_{AH}^{ 5D}=\mathbf{S}(v)\pm \sqrt{M(v)+\mathbf{S}(v)^2}$ with $\mathbf{S}(v)=S(v)/3$ and at $r_{AH}^{6D} = \eta(v)^{1/3}/6+S(v)/ \eta(v)^{1/3}$ with $\eta(v):=72M(v) + 6 \sqrt{144M(v)^2-6S(v)^3}$ for the 6D case. Thus in order to have horizon in $6D$ or $r_{AH}^{6D}$ to be real valued, we must have $144m(v)^2 \geq 6S(v)^3$, otherwise we have no horizons and only a naked singularity.
\subparagraph{EGB case:}
From (\ref{thgb}) it is clear that the apparent horizon is the solution
of
\begin{equation}\label{aegb}
1+\frac{r^{2}}{2{\alpha}}\left(  1\pm
\sqrt{1+\frac{4{\alpha}M(v)}{(N-2)r^{N-1}} + \frac{8 {\alpha}S(v)}{(N-2)r^{N-2}}} \right)=0.
\end{equation}
It is clear that  (\ref{aegb}) may not obviously admit simple closed solutions.  However, it is easy to get solutions in the 5D and 6D cases as
\begin{eqnarray}\label{aegb}
 r_{AH}^{5DGB} & = & \mathbf{S}(v) \pm \sqrt{M(v) - {\alpha} + \mathbf{S}(v)^2 },\nonumber \\
 r_{AH}^{6DGB}& = & \frac{\xi(v)^{1/3}}{6}+\frac{\left(S(v) - 2\alpha\right)}{\xi(v)^{1/3}} \nonumber \\
\xi(v) & = & 27 M(v) + 3 \sqrt{\gamma(v)}.
\end{eqnarray}
with $\gamma(v) = 18 M(v)^2-24 S(v)^3+144 \alpha S(v)^2-288\alpha^2S(v).$
In the relativistic limit $\alpha \rightarrow 0$, with $S(v)=0$,  $r_{AH}^{5D} \rightarrow \sqrt{M(v)}$.
Further, the solutions have the right limit when $S(v)=0$ and/or $\alpha=0$ and for the $5D$ case we regain  the solution obtained in \cite{dwd}.
We see that $g_{vv}(v,r_{AH})  = 0$  at the apparent horizons implies that  they are
 also timelike surfaces.   It is clear that presence of the  coupling constant
of  the Gauss-Bonnet term $\alpha$ and string clouds in background produce a change in the location of the apparent horizon.  Such a change could have a significant effect in the dynamical evolution of the black hole horizon.  For an outgoing null geodesic $r = r_{AH}$, is given by
\begin{equation}\label{rd}
 \dot{r}=\frac{dr}{dv} = \frac{1}{2}f(v,r).
\end{equation}
For the EGB case, differentiating (\ref{rd}) with respect to $v$, we obtain
\begin{align}\label{rdd}
& \ddot{r}= \frac{d^2r}{dv^2} = \frac{r\dot{r} \;\Sigma(v,r)}{4\alpha}  + \frac{4 \alpha r^2}{\Sigma(v,r)} \Big[ \frac{L_M}{(N-2)r^{(N-1)}}  \nonumber \\ & +  \frac{L_S}{(N-2)r^{(N-2)}} + \frac{(N-1)
M(v)\dot{r}}{(N-2)r^N}+ \frac{2S(v) \dot{r}}{r^{(N-1)}} \Big],
\end{align}
with
\begin{equation}
\Sigma(v,r)= \left(  1\pm
\sqrt{1+\frac{4\tilde{\alpha}_{2}M(v)}{(N-2)r^{N-1}} + \frac{8 \tilde{\alpha_2}S(v)}{(N-2)r^{N-2}}} \right).
\end{equation}
At the timelike surface or apparent horizon $r = r_{AH}$, $\dot{r} = 0$
and $\ddot{r}> 0$ for $L_M>0$ and $L_S >0$.  Hence the photon will escape from the $r=r_{AH}$ and reach an arbitrary large distance, which confirms that the surface $r = r_{AH}$ is an apparent horizon not an event horizon.  Thus the apparent horizon is the outermost maximally trapped surface for an outgoing photon.
On the other hand, the event horizon  is a null three-surface which is the locus of outgoing future-directed null geodesic rays  that never manage to reach arbitrarily large distances from
the black hole and are different from the apparent horizon for radiating black holes. They are determined via the Raychaudhuri equation; it can be seen to be equivalent to the requirement that ${d^2\Theta}/{dv^2} \approx 0$ to $O(L_M,\;L_S)$ \cite{jy,rm,mrm}.  It can be shown that the expressions for event horizons are the same as  apparent horizons, but  $M(v)$ and $S(v)$ replaced by $M^*(v)$ and $S^*(v)$, where
\begin{equation}
M^*(v) \approx M(v) - \frac{L_M}{\kappa},\;
S^*(v) \approx S(v) - \frac{L_S}{\kappa}.
\end{equation}
Therefore a new region ($r_{EH} <r<r_{AH}$), called  the {{\it quantum ergosphere} \cite{jy}} exists for radiating black holes, which is absent in the static black holes. In this region photons are locally trapped but, being outside the event horizon, they can cross the apparent horizon at a
later time and propagate to infinity. It turns out that because of evaporation apparent and event horizons, coincide for the static Schwarzschild solution.
Finally, the Hawking temperature near the apparent horizon can be obtained through the relation $T=\kappa/2\pi$.

\section{CONCLUSION}
Lovelock theory is a natural extension of Einstein's GR to higher dimensions
in which the first, second and third order terms correspond, respectively,
to  the higher dimensional GR and EGB gravity and third order Lovelock gravity.
In this paper, we have obtained exact radiating black holes in the background of a cloud of strings in arbitrary $N$ dimensions in these theories.  Thus we have  explicit nonstatic radiating black hole solutions in these  theories. The black holes are characterized by mass the $ M(v) $, the string cloud function $S(v)$ and the parameters $\alpha$ and $\beta$, and the solutions are asymptotically dS (AdS).
We have
used the solutions to discuss the consequence of the GB term and string clouds on the structure and location of the horizons for  radiating
black holes. By defining all kinematical parameter in terms  of null vectors, and using the definition suggested by York \cite{jy}, we have investigated the structure and locations of horizons for both the GR and EGB cases. The apparent horizons are obtained exactly and event horizons are obtained to
first order in the luminosity using the  method developed by York \cite{jy}.
We have shown that a radiating black hole  has three important
horizon-like loci that full characterize its structure, viz. apparent horizon, event horizon
and a timelike limit surface; we have the relationship of the three surfaces $r_{EH} < r_{AH} = r_{TLS}$
and the region between the apparent horizon and event horizons  is defined as the
\emph{quantum ergosphere} \cite{jy}. The presence of the coupling constant of
the Gauss-Bonnet terms and string clouds produce a change in the location of
these horizons. Such a change could have a significant
effect in the dynamical evolution of these horizons.

The static black hole
solutions in  Eddington-€"Finkelstein coordinates, in both HD and in 4D, can be
recovered by setting $M(v) = M, \; S(v) = S$, with $ M $ and $ S $ as
constants in which case $f(v,r) \rightarrow f(r)$.
In the static limit, we can  obtain from the metric (\ref{megb}),
the usual spherically symmetric form
\begin{equation}
ds^2 = -f(r)\; dt^2 + \frac{dr^2}{f(r)} + r^2 d \left(\Omega_{N-2}\right)^2,
\end{equation}
by the coordinate transformation
\begin{equation}
dv = A(r)^{-1} \left( dt + \epsilon \frac{dr}{f(r)} \right).
\end{equation}
In the case of spherical symmetry, even when $f(r)$ is replaced by
$f(t,r)$, we can cast the metric in the form (\ref{megb})
\cite{visser}.

A rigorous formulation and proof for the cosmic censorship conjecture  is far from our reach.  Hence, examples or counterexamples
remain the only tool to study the various aspects of this important conjecture.  However, the lack of exact solutions,
suitable to study gravitational collapse from the viewpoint of cosmic censorship  makes progress very difficult.  As a consequence,  we are far
from completely understanding even the simple case of
spherical symmetry. The solutions presented here are dynamical which are suitable for such studies and  can be useful to
get insights into more general gravitational collapse situations,
and, in general, a better understanding of the conjecture that may help to formulate it
in precise mathematical form.

\section*{Acknowledgments}
S.G.G.  thanks the
University Grant Commission (UGC) for the major research project
grant F. NO. 39-459/2010 (SR). S.D.M. acknowledges that this work is
based upon research supported by the South African Research Chair
Initiative of the Department of Science and Technology and the
National Research Foundation.

\appendix*

\section{Radiating black hole in third order Lovelock theory}
In the present section, we proceed to consider third order Lovelock theory, i.e., we choose both $\alpha$ and $\beta$ to be nonzero.   If $N\geq7$, with both $\alpha$ $\beta$ nonzero, the role of the third order Lovelock term as well as the second order Gauss-Bonnet term will be nontrivial to the gravitational dynamical equations. Hence, the radiating black holes  governed by the Eq. (\ref{f-egb}),  can be  integrated in arbitrary dimension $N\geq7$ and solution can be expressed as:
\begin{eqnarray}\label{solll}
f(v,r) & = & 1 + \frac{\alpha\; r^2}{\beta} -  \frac{2 r^{2N+4}(\beta-\alpha^2)(N-2)}{\beta r^N \Delta(v,r)^{1/3}} \nonumber \\ & & + \frac{\Delta(v,r)^{1/3}}{2 \beta r^N (N-2)},
\end{eqnarray}
with
\begin{align}
\Delta(v,r)&=4 (N-2)^2 \beta r^N \sqrt{\Xi} + \left( \beta -  \frac{2\alpha^2}{3}\right) (N-2)^{3} \alpha r^{3N+6}\nonumber \\ &   - 2 r^{2N+7}\beta^2 \chi(v,r) (N-2)^{2}, \nonumber \\
\Xi(v,r) & = 36 \chi(v,r) r^{2 N + 14} \beta^2 \nonumber \\&  + 4 r^N \beta (N-2)^2 r^{3 N + 12} \nonumber \\& - 3 \alpha^2 (N-2)^2 r^{4N+12} - 36  \left( \beta -  \frac{2\alpha^2}{3}\right) (N-2) \alpha r^{3 N + 13}  \chi(v,r),\nonumber \\
\chi(v,r)& = S(v) r - \frac{M(v)}{6}.
\end{align}
The special, but  interesting case $\beta =  {2\alpha^2}/{3}$, immensely simplifies the above solutions as
\begin{equation}\label{solllb}
f(v,r)= 1 + \frac{3 r^2}{2 \alpha} +
\left(\frac{9}{8}\right)^{\frac{1}{3}}\frac{r^{2N+4}
(N-2)}{r^N\Delta(v,r)^{1/3}}+
\frac{3^{\frac{1}{3}}\Delta(v,r)^{1/3}}{2 \alpha^2 r^N (N-2)},
\end{equation}
where
\begin{align}
& \Delta = {\alpha}^{2}\left( N-2 \right) ^{2} \Big[ \sqrt {8}\sqrt {3}\delta {r}^{N} + 12\,{\alpha}^{2}{r}^{2\,N+7} \chi(v,r) \Big] , \nonumber \\
& \delta = \sqrt {{\alpha}^{2} \left( {r}^{
N} \left( N-2 \right) ^{2}{r}^{12+3\,N}-{\frac {9}{8}}\,{r}^{12+4\,N}
 \left( N-2 \right) ^{2}+ \tau(v,r)
 \right) },
\end{align}
with $\tau(v,r) = 6\, \chi(v,r)^{2}{r}^{N}{r}^{N+14}{\alpha}^{2}$
and $\chi(v,r)= S(v) r - {M(v)}/{6}$. Another interesting case arises when $\beta\neq0$ but $\alpha=0$, which can be identified as Einstein-Lovelock theory.  In this case the solution further simplifies to
\begin{align}
f(v,r) = 1 + \frac{\Delta^{1/3}}{2\beta r^N(N-2)} - \frac{2 r^{(N+4)}(N-2)}{\Delta^{1/3}},
\end{align}
with
\begin{align}
& \Delta = -4 r^{7}\left(6 S(v) r + M(v)\right)+ (N-2)^2 \beta^2 r^{(2N+6)} \nonumber \\
& \times \frac{4}{\sqrt{\beta}}   \sqrt{4 (N-2)^2 r^{(2N)} + \beta
r^{2} \left(6S(v)r +  M(v) \right)^2}.
\end{align}
Thus we obtain a kind of radiating Vaidya black hole spacetime, with clouds of string, in  third order Lovelock gravity by solving Eq.~(\ref{f-egb}). When the dimension of spacetime is five, that is $N=5$, and $M(v)$, $S(v)$ constants, the solution reduces to the one reported by some authors in Ref. \cite{he}. In $N=4$ and $S(v)=0$, the solution is just the familiar Vaidya solution \cite{pc} of general relativity.

\end{document}